\documentclass[3p,times,twocolumn]{elsarticle}

\usepackage{amssymb}

\usepackage[figuresright]{rotating}
\usepackage{epsfig}

\begin{document}

\begin{frontmatter}


\title{Progress of the MICE experiment at RAL}
\author{M. Bonesini\fnref{lab1}}
\fntext[lab1]{On behalf of the MICE collaboration}
 \address{Sezione INFN Milano Bicocca\\Dipartimento di Fisica G. Occhialini,
  Universit\`a di Milano Bicocca, Milano}






\begin{abstract}
The international Muon Ionization Cooling Experiment (MICE) will perform a
systematic investigation of ionization cooling of a muon beam.
The demonstration comprises one cell of the US  Neutrino Factory Study II
cooling channel. Results obtained on the construction of the beamline 
and its instrumentation (STEP I) will be reviewed, together with progress 
towards final measurements of ionization cooling (STEP IV and VI). 
\end{abstract}

\begin{keyword}
Ionization cooling, neutrino beams, neutrino factory, muon collider


\end{keyword}

\end{frontmatter}


The proposed Neutrino Factory (NF)~\cite{Koshkarev,Geer} is a muon
storage ring with long straight
sections, where decaying muons produce collimated neutrino beams of high
intensity and well defined
composition, with no uncertainties in the spectrum and flux
from hadronic production~\cite{Bonesini}.
The cooling of muons
will increase the NF performance and reduce the
muon beam emittance up to a factor 2.4
(as described in reference \cite{Choubey} with a cooling section 75 m long)
\footnote{By contrast, cooling requirements for a Muon Collider are much
more demanding, requiring cooling factors up to $10^6$}.
A NF will be the most efficient tool to probe neutrino
oscillations and observe CP violation in the lepton sector.

The MICE experiment \cite{mice} at RAL aims at a
systematic study of a section of the  cooling channel of the 
proposed US NF Study 2~\cite{US2a}, 
attaining a $10 \%$ emittance reduction for a $6  \pi \cdot$mm rad beam.
The 5.5 m long cooling section consists of three liquid hydrogen absorbers,
inside focus coil modules (AFC) and eight 201 MHz RF cavities surrounded by lattice solenoids.
\begin{figure*}[thb]
\vskip -0.6cm
\begin{center}
\includegraphics[width=0.85\linewidth]{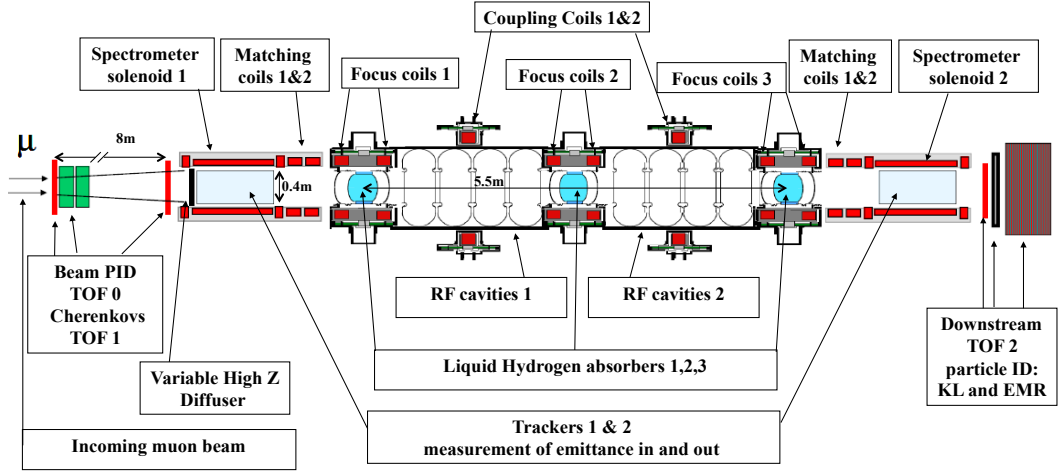}
\end{center}
\vskip -4.0cm
\caption{View of the MICE experiment at RAL.
The muon beam from ISIS enters from the left. The cooling channel
is put between two magnetic tracking spectrometers and two TOF stations
(TOF1 and TOF2) that measure particle parameters.}
\label{fig:mice}
\end{figure*}
The secondary muon beam from ISIS (140-240 MeV/c central momentum, 
tunable between $3-10 \pi \cdot  $ mm rad  input emittance) enters
the MICE cooling section after a diffuser of adjustable thickness.
Muons originate from $\pi$
decay inside a 5 m long superconducting (SC) solenoid upstream of 
the first timing detector (TOF0). 
As conventional emittance measurement techniques, based on profile
monitors,  reach
 barely a $\sim 10 \%$
precision, a novel method based on single particle measurements
has been proposed.
For each particle $ x,y,t$,$p_x$,$p_y$, $E$ coordinates
are measured before and after the cooling section.
In this way,  the
input and output beam emittances may be  determined  with a precision
up to $0.1 \%$, that allows a sensible extrapolation of the results 
to the full cooling channel.
The experiment will be done in several steps, of which the first one (STEP I)
is the characterization of the beamline~\cite{bogomilov}.
\section{MICE STEPI }

The driving design criteria for MICE beamline detectors are robustness,
in particular of the trackers, to sustain the severe
background conditions near the RF cavities and redundancy in particle
identification (PID)
to keep beam contaminations ($e, \pi$) well below $1 \%$ and
reduce systematics on the emittance measurements.

PID is obtained upstream of the first tracking solenoid by two
time of flight (TOF) stations (TOF0/TOF1) \cite{yordan} and two threshold Cerenkov counters (CKOVa/CKOVb)
\cite{ckv},
that will provide $\pi/\mu$ separation up to 365 MeV/c. 

\noindent
Downstream the PID is obtained instead using an additional TOF station (TOF2)
and a calorimeter, made of two detectors (KL and EMR),
to separate muons from decay electrons and undecayed pions.
While EMR determines precisely the muon 
momentum by range measurement, KL acts as an active pre-shower to
tag electrons.
All TOF detectors are used to determine the time coordinate ($t$) 
in the measurement of the emittance.

\begin{figure}[hbt]
\begin{center}
\includegraphics[width=\linewidth]{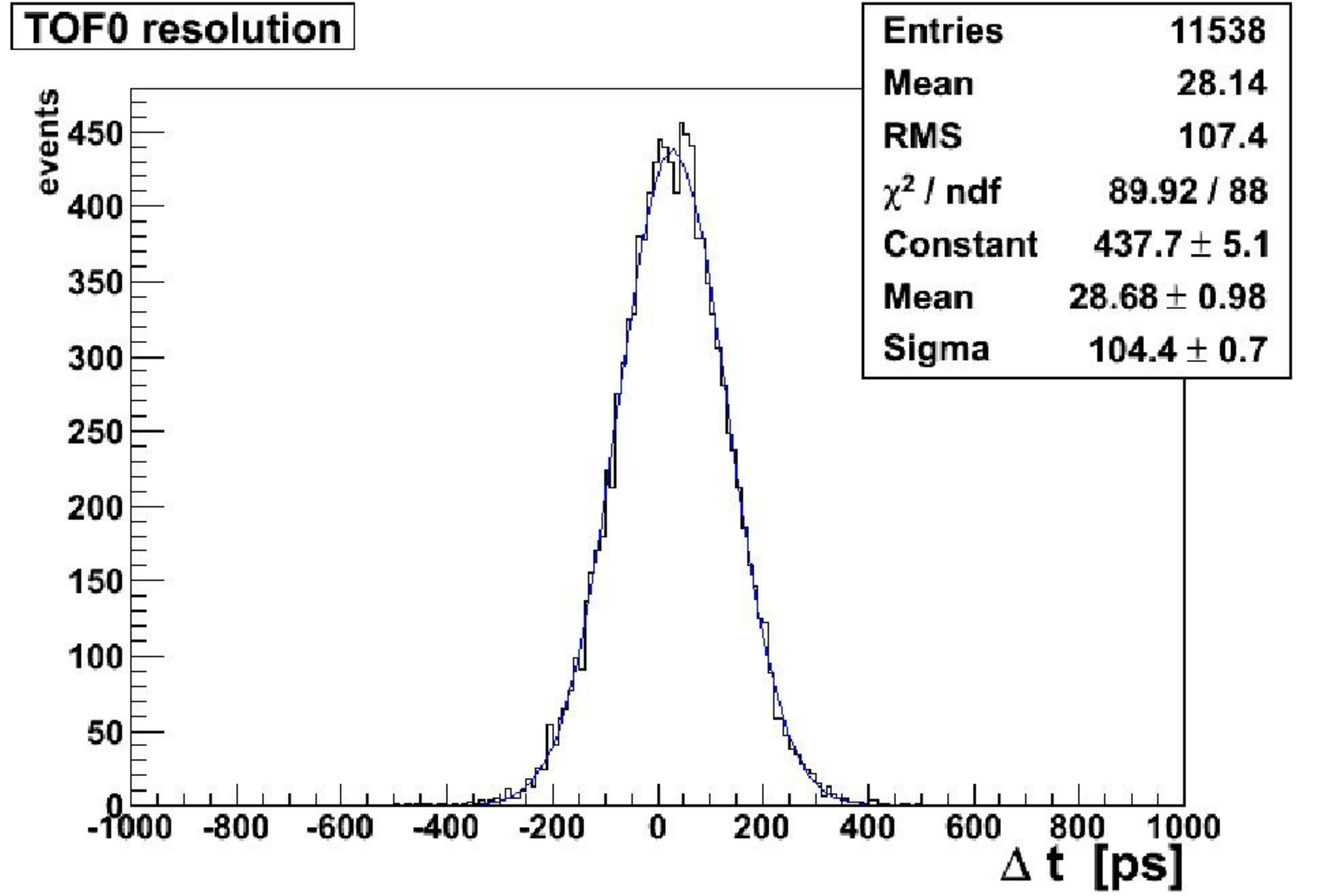}
\end{center}
\vskip -0.5 cm
\caption{ Time difference $\Delta t_{xy}$ between $x/y$  slabs in TOF0.}
\label{fig:TOF2}
\end{figure}

After time-walk corrections and 
calibration  with impinging beam particles~\cite{yordan},
the TOF detector timing resolution can be measured by using the time
difference $\Delta t_{xy}$  between the vertical and horizontal slabs in 
the same station (see  figure \ref{fig:TOF2} for an example).
The obtained resolution on the difference  $\sigma_{xy}\sim 100 \ ps$
 translates into $\sim 50 \ ps$
 time resolution for a full TOF detector with crossed $x/y$ slabs.

The characterization of the MICE beam line has been done mainly using the TOF
detectors. 
\begin{figure*}[hbt!]
\vspace{-4.2cm}
\begin{center}
\includegraphics[width=0.64\linewidth]{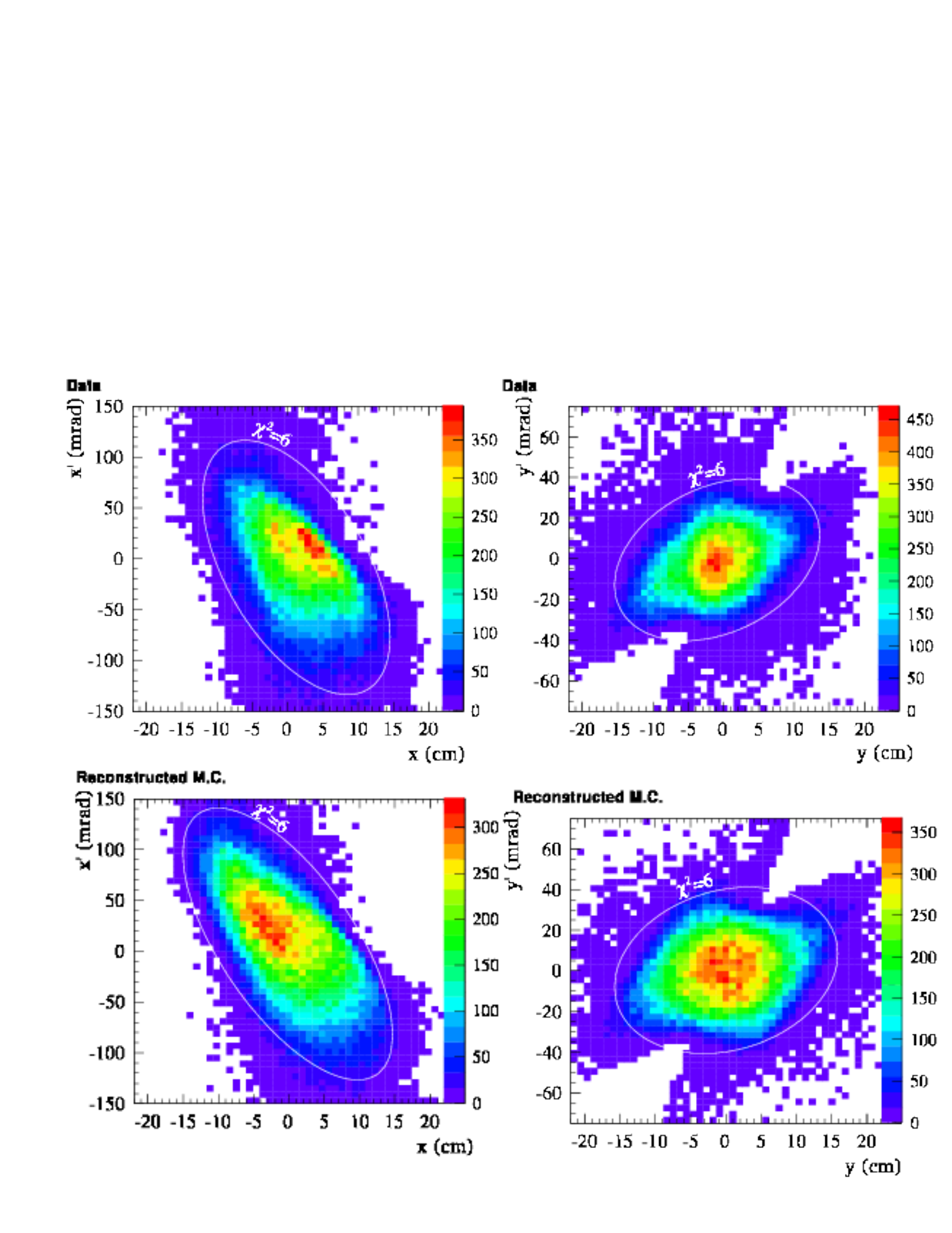}
\end{center}
\vskip -1.2cm
\caption{ 
Reconstructed data (top) and simulation (bottom) horizontal ($x$)
and vertical ($y$) trace plots at TOF1 
for the baseline $\mu^{-}$ beam $\epsilon=6 \pi$ mm and $p_z=200$ MeV/c.}
\label{fig:trace}
\end{figure*}

Waiting for the delivery of 
the tracking solenoids, the beam emittance was preliminary 
measured, in MICE STEP I,  with the TOF detectors only.
TOF detectors were used to derive  the $x,y,x',y'$
information for each particle and measure $p_z$ from the time-of-flight
between TOF0 and TOF1.       
Once the initial and final $x,y$ particle coordinates are measured, 
the muon track through the present MICE channel 
is estimated as $(u_1,u'_1)={\bold M} (p_z) \cdot (u_0,u'_0) $, 
with $\bold{M}(p_z)$ transfer matrix and $u=x$ or $y$. 
The muon momentum $p_z$ is initially estimated via the formula:
$ p_z=E \cdot (s/ \Delta t) $,
with $s$ track length and $ \Delta t$ time-of-flight between TOF0 and TOF1.
With a separation of $\sim 7.7$ m between TOF0 and TOF1, $p_z$ may be
measured with a resolution $\sigma_{p_z} = (E^2/m^2)\sigma_t/\Delta t$.
For  the  baseline beam, with $\epsilon=6 \pi$ mm 
and $p_z=200$ MeV/c, this corresponds to about $1 \%$, giving a comparable
resolution on the transfer matrix $\bold{M}$.  
An iterative procedure, based on the transfer matrix, is then used to 
recompute $s$ and $p_z$. This procedure correct a  track length bias
($\sim 5$ MeV/c) on $p_z$ reducing it to less than 1 MeV/c. 

At this point $x',y'$ are evaluated from the
initial and final muon positions. 
The horizontal and vertical trace space distributions $(x,x')$ and
$(y,y')$ are then plotted and the transverse emittances $\epsilon_x$ and
$\epsilon_y$ may be computed, giving an estimation
of the normalized emittance via the formula:
$ \epsilon_{N} \simeq (p_z/m) \sqrt{\epsilon_x \cdot \epsilon_y}$. 
Figure 3  shows the trace plots 
for the MICE $\mu^{-}$ baseline beam, with $\epsilon=6 \pi$ mm 
and $p_z=200$ MeV/c, for both experimental data and MC simulation. Even if
the agreement is not perfect, the beam occupies the expected regions in 
trace space. All beams show an RMS beam size of the order of 5-7 cm. 
This method allows to measure  beam emittances at the few per-cent level. 
\section{Progress towards STEP IV and VI of MICE}
In STEP IV, with the full beam instrumentation system available (including the two
trackers \cite{Ellis})  and one Absorber Focus
Coil (AFC) in between, it will be possible to measure  the input
and output beam emittances and evaluate the ionization process with the full
design precision. 
The AFC focus coil, shown in figure 4, provides the guiding
magnetic field in the liquid hydrogen ($LH_2$) absorber. Just delivered
to RAL, the system is under test with the $LH_2$ system for full integration
inside MICE STEP IV.
\begin{figure}[hbt!]
\begin{center}
\includegraphics[width=0.35\linewidth]{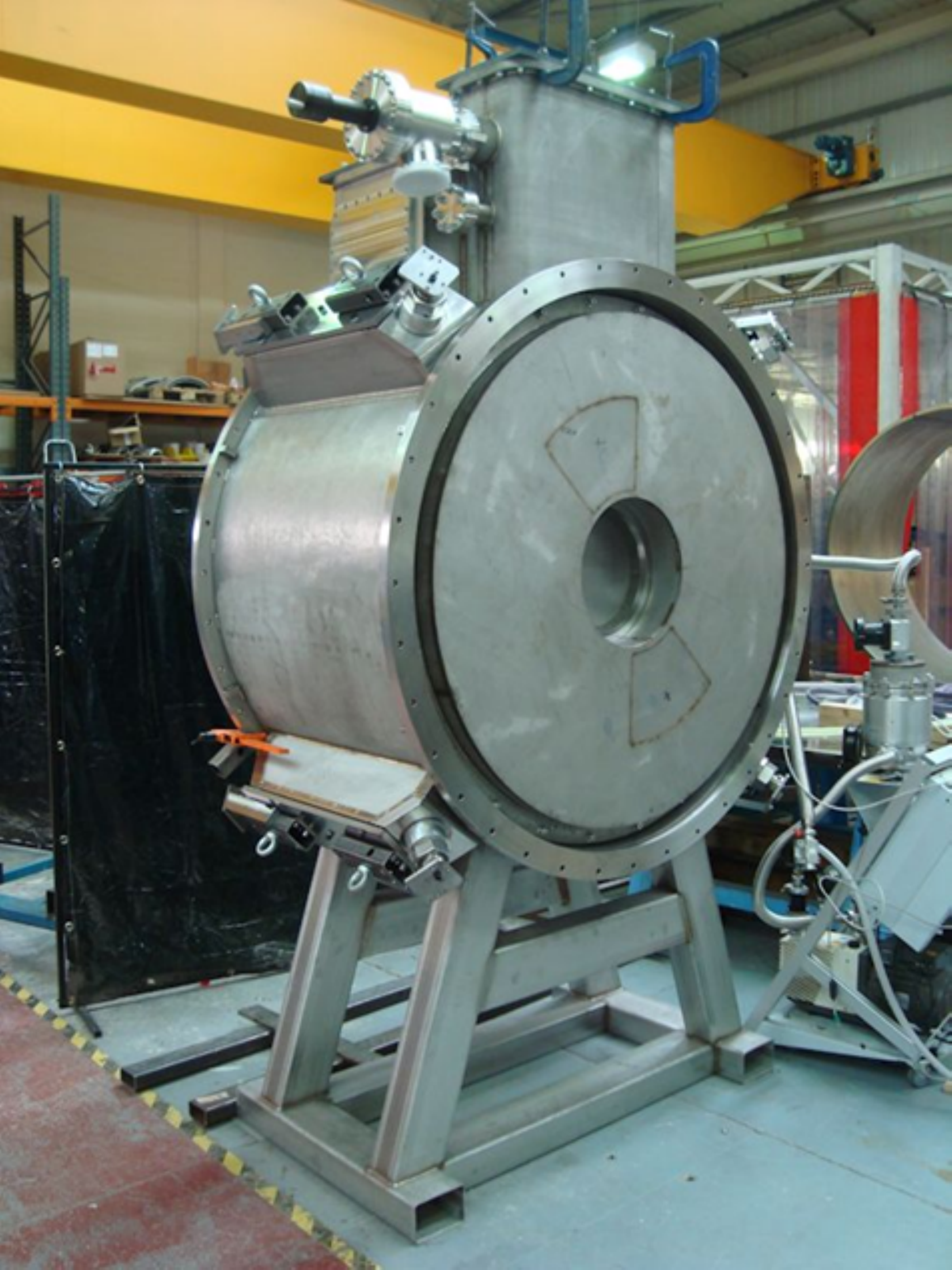}
\includegraphics[width=0.63\linewidth]{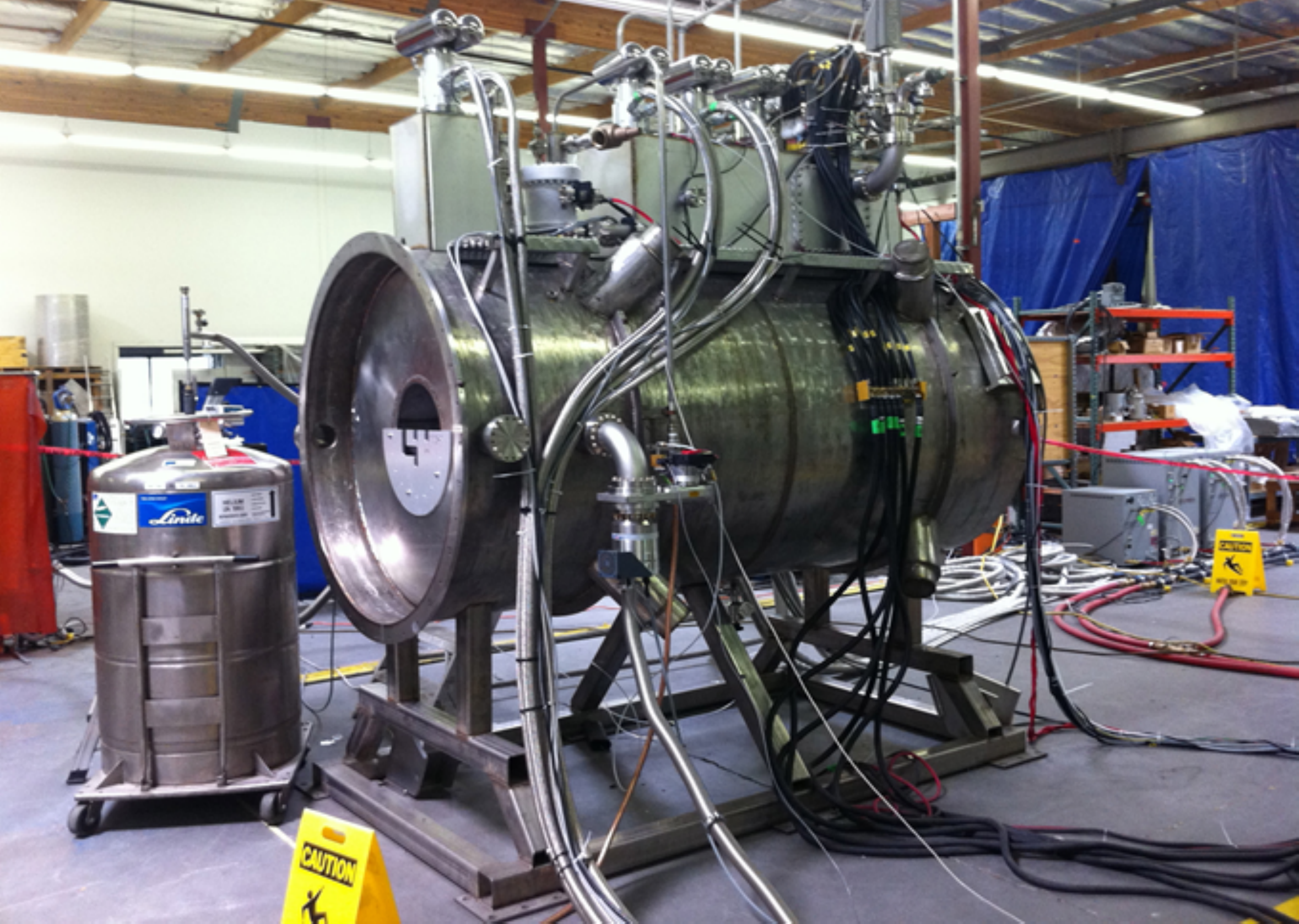}
\caption{Left panel: Absorber Focus Coil (AFC). Right panel: SC solenoid 
under training at Wang NMR.}
\end{center}
\label{fig:AFC}
\end{figure}
The SC tracking solenoids provide a guiding magnetic field in the fiber 
tracker. Figure 4 shows the training of the first one at Wang
NMR manufacturer. Modifications were implemented to reduce the heat leak to the cold mass,
increase the available cooling power via cryocoolers and replace  LTS
and HTS faulty leads.   
For STEP VI, two RFCC modules will be installed to compensate
for the longitudinal momentum loss and two additional AFC will be installed
to increase the ionization effect, responsible of the muon cooling. 
Each RFCC module consists of four 201 MHz RF cavities and one coupling coil
(CC), that will provide the guiding magnetic field inside the cavities.
Construction of STEP IV components is almost complete, while 
STEP VI has 2016 as target date.   

\end{document}